\documentclass[manuscript]{aastex}
\usepackage{CJK}

\slugcomment{2012, ApJ, 760, 115}

\shorttitle{Multipolar PNe 3D Geometry}
\shortauthors{Chong et al.}

\begin{document}

\title{Multipolar Planetary Nebulae:\\
Not as Geometrically Diversified as Thought}
\begin{CJK*}{Bg5}{bsmi}

\author{S.-N. Chong (²ø«ä¹ç)\altaffilmark{1}}
\affil{Graduate School of Science and Engineering, Kagoshima University, 1-21-35 Korimoto, Kagoshima 890-0065, Japan}
\email{chongsnco@gmail.com}

\author{Sun Kwok (³¢·s)\altaffilmark{2}}
\affil{Department of Physics, The University of Hong Kong, Pokfulam Road, Hong Kong}
 \email{sunkwok@hku.hk}

\author{H. Imai \altaffilmark{1}}

\author{D. Tafoya\altaffilmark{3}}
\affil{Onsala Space Observatory, SE-439 92 Onsala, Sweden}

\and
\author{J. Chibueze\altaffilmark{1}}

\begin{abstract}

Planetary nebulae (PNe) have diverse morphological shapes, including point-symmetric and multipolar structures.  Many PNe also have complicated internal structures such as torus, lobes, knots, and ansae.  A complete accounting of all the morphological structures through physical models is difficult.  A first step toward such an understanding is to derive the true three-dimensional structure of the nebulae.  In this paper, we show that a multipolar nebula with three pairs of lobes can explain many of such features, if orientation and sensitivity effects are taken into account.  Using only six parameters $-$ the inclination and position angles of each pair $-$ we are able to simulate the observed images of 20 PNe with complex structures.  We suggest that the multipolar structure is an intrinsic structure of PNe and the statistics of multipolar PNe has been severely underestimated in the past.
\end{abstract}


\maketitle
\end{CJK*}

\section{Introduction} 

One of the fascinating aspects of planetary nebulae (PNe) is their diverse morphological shapes.  Since the original morphological classification by \citet{cur18}, many attempts have been made to classify the apparent morphology of PNe \citep{bal87, sta93, aaq96, man96, par06} using descriptive terms such as ``round'', ``elliptical'', and ``bipolar''.  As the sensitivity of imaging techniques improved, increasing numbers of PNe were found to possess more than one pair of bipolar lobes.      In early literature, the terms ``twofold bipolarity", ``overlapping bipolar structures" and ``bi-butterfly" were used to describe these objects \citep{pas90a, pas90b}.   \citet{man96} introduced a new morphological class named ``quadrupolar": lobe pairs symmetric about two distinct axes.
In general, the term ``multipolar" refers to morphologies with at least two axes of symmetry.
It is sometimes referred to as ``polypolar" \citep{lop98}.
The multipolar morphology is not only found in PNe, but is also observed in earlier phases of post-asymptotic giant branch (post-AGB) objects including pre-planetary nebulae (pPNe).

Deeper imaging of PNe has revealed that some well-known PNe are actually multipolar.  For example, both NGC~2440 and NGC~6072 were classified as bipolar by \citet{cor95}.
However, detailed spatial and kinematical studies of optical lines by \citet{lop98} confirmed the presence of two pairs of lobes in NGC~2440, also identified in CO observations by \citet{wan08}.
Similarly, NGC~6072 appears to be an elliptical shell with faint extensions in the Digital Sky Survey optical image, but the \textit{Spitzer Space Telescope} Infrared Array Camera (IRAC) and Canada$-$France$-$Hawaii Telescope (CFHT) molecular hydrogen (H$_2$) images revealed a complicated structure with three pairs of lobes and an elliptical ring \citep{kwo10}.  NGC 7293 (the Helix Nebula) is known to have at least two bipolar outflows \citep{mea08}.
NGC~6853, the Dumbbell Nebula, shows a pair of cones in addition to the well known dumbbell-shaped main nebula in the IRAC and CFHT H$_2$ images \citep{kwo08}.
In these examples, it can be seen that deeper exposures and observations in other wavelengths can result in the identification of multipolar lobes.

Statistical results of the PNe morphology also show an increasing fraction of the multipolar class.
In early classifications, there were no multipolar categories, and such objects might have been included in the ``point-symmetric" or ``irregular" classes \citep[e.g.,][]{bal87, sch93b, cor95}.
In more recent results, from \textit{Hubble Space Telescope (HST)} images, 12\% of 140 post-AGB objects and 20\%  of 119 young PNe were classified as multipolar by \citet{man11} and \citet{sah11}, respectively.
The trend may still be growing.


Efforts have been made to construct three-dimensional (3D) models of individual multipolar PNe:
e.g. NGC~6644 \citep{hsi10}, Hb~5 \citep{gar11}, and J~320 \citep{har04}.
However, as these models are tailored-made for each PN, the similarities and differences among objects are unclear, which hinders our understanding of the origin of multipolar morphology.
In order to explain the presence of multiple outflow axes, one has to introduce additional hypotheses, such as precessing jets in the short transition period between the spherically symmetric AGB phase and the multipolar post-AGB phase.  Bipolar, rotating episodic jets provide direct evidence of precession motions as observed in Fleming~1 \citep{lop93} and KjPn~8 \citep{lop95}.
It is suggested that a rotating jet can account for the creation of continuous point-symmetric features at different distances from the geometric center, e.g., IRAS~16585-2145 \citep{gue08} and IRAS~17028-1004 \citep{cor11}.   Moreover, precessing motions traced by water maser spots in early stages \citep[e.g.,][]{yun11} may bring out an indirect way of proof.  It is still under debate whether the multiple lobes are formed simultaneously or episodically \citep{sah02}. They may involve totally different physical processes.

Before starting to establish the theories, the first step should be to determine the actual 3D structure, rather than applying only the projected two-dimensional (2D) images of these PNe.
Based on the 3D model, one can estimate the kinematic timescale in each outflow direction to verify whether they were produced at the same time.   It has been pointed out that the current classification schemes do not reflect the true 3D physical structure of the nebulae as the observed images suffer from effects of sensitivity dependence, atomic or molecular species dependence, and projection effects \citep{kwok2010}.   
As a first step toward understanding the true 3D structure of PNe, it would be useful to  build a unified 3D model to see how far one can go to reproduce the observed 2D images of individual objects.  By making this attempt, we are trying to address the question of whether ``each PN is unique'' or whether they are just different manifestations of a unified structure.   

In this paper, we explore answers to this question by beginning with a simple model of a PN with multiple lobes that all have the same shape, kinematic timescales and outflow velocities.
A sample selection of real PN images to be compared is in the next section, and the model is described in detail in Section~\ref{model}.
Section~\ref{results} presents the projected images of the model in different parameters and a comparison with real PN images.
Discussions including the meanings of parameters and the formation mechanisms can be found in Section~\ref{discuss}.

\section{Sample Selection} 
To compare the modeled images with real observed ones, we searched through the literature \citep[][and references therein]{sah11, uet07, gue08, har04} for objects which have been observed in H$\alpha$ with the Wide-Field Planetary Camera 2 of the \textit{HST}.
Emission in the recombination line H$\alpha$ clearly indicates that the sources are PNe rather than pPNe in which the central stars are not hot enough for photoionization.
Most of the objects are called young PNe according to their [O~{\sc iii}]$\lambda$5007/H$\alpha$ flux ratio of less than about unity \citep{sah11}.
H$\alpha$ emission, as the most commonly observed line of PNe, provides a larger pool for sample selection in this paper and future projects.
The sample contains a total of 20 PN images retrieved from the Canadian Astronomy Data Center archive (details are listed in Table~\ref{tab1}).
These 20 objects are selected due to their resemblance to the projected images of our model.
Therefore, we do not claim that they represent the whole PN set.

\section{The Model} \label{model} 

Our model images are constructed using the 3D modeling software SHAPE \citep{ste11}.
SHAPE is an interactive tool for morpho-kinematic modeling and image reconstruction.
Customized objects can be constructed by starting with built-in basic geometrical shapes and adding modifiers to them.

We begin with a simple model where the nebula consists of three pairs of identical lobes.  
We first address the issue of projection effects on the lobes as observed from different orientations. All other parameters such as the density profiles are fixed and the inclination angle $i$ and position angle (PA) of each pair are the only variable parameters.
$i$ is taken to be the angle between the lobe axis and the line of sight (LOS).
As a result, there are only six independent parameters.  From these six parameters, the separation angle $\theta$ between any two pairs of lobes can be calculated from the inner product.  There are three values of $\theta$ for three pairs of lobes.  The lobes are hollow inside with evenly distributed density within the ``walls" of the lobes.   The brightness is proportional to the square of the column density based on the fact that the intensity of H recombination lines is directly proportional to the square of the electron density \citep{kwok07}.

\section{Results} \label{results} 

\subsection{Apparent Morphologies}\label{appmor}

Figure~\ref{figrdm} shows the projected images of 100 randomly generated combinations of the three-paired model.
Each pair of lobes in each image has an independent random orientation in the 3D space: PA is a random multiple of five between $-$175 and +180 inclusive (i.e., $-$175, $-$170, $-$165, ..., 0, ..., +180); $i$ is a random integer between 0 and 90 inclusive with a weight proportional to $\sin i$, as the differential solid angle in polar coordinates is $(\sin i\ d(i)\ d$(PA)). This expression indicates that a pair of lobes has a higher chance to be viewed edge-on than pole-on.
The values of $i$ and PA are listed in Table~\ref{tblrdm}. 
It can be seen that although the true sizes of lobes are the same, the apparent length changes with the viewing angle.  The combination of three pairs greatly increases the variance.
In some special cases, the multipolar nature of the nebula is not obvious from the projected images, for example, when projections of two or more pairs of lobes are aligned along the same direction so that the lobes overlap with each other as seen by the observer.
If one pair is viewed nearly pole-on or slightly tilted in the equatorial direction, then it may be wrongly interpreted as a feature in the equatorial direction or be labeled as a torus.

From this simulation, we can estimate the fraction of apparent morphologies created as the result of random orientation.
The visual appearance of these images can be classified into the following classes:

\begin{itemize}
\item{Class 0: a round or elliptical appearance and no obvious lobes. The fraction belonging to this class is 0\%.}
\item{Class 1: one pair of lobes, with or without a torus-like structure.  The fraction belonging to this class is 5($\pm4$)\%.}
\item{Class 2: two pairs of lobes, with similar or different lengths.  The fraction belonging to this class is 46($\pm5$)\%.}
\item{Class 3: three pairs of lobes.  The fraction belonging to this class is 49($\pm8$)\%.}
\end{itemize}
We have adopted some ``usual customs" of morphological classification of PNe: if two pairs of lobes look nearly perpendicular and one of them is significantly shorter, then the short pair is usually identified as the torus; shorter lobes overlapped with longer ones are called inner lobes rather than extra pairs of lobes.
The classification of each image has been done by four individuals independently, and the overall results are listed in Table~\ref{tblrdm}.
The percentages above are from the overall results, and errors correspond to variations among individuals, which are within expectation and actually reflect the real situation of previous classification works.

This exercise shows that by assuming the number of pairs is at most three, only half of the PNe with three pairs of lobes will be correctly identified. The other half will have the number of lobes underestimated, and 5\% of them will even be misclassified as single-paired, or more commonly called, bipolar.
In other words, the true number of those with three pairs of lobes should be $\sim$2 times the number of visually identified ones.
Denoting $n(a,b)$ as the number of PNe with $a$-pairs that are intrinsic but only with $b$-pairs that are observed, the above statement can be expressed as \[ \sum_{m=0}^3 n(3,m) \sim 2n(3,3). \]
In general, the true number of those with $x$-pairs intrinsic should be 
\begin{eqnarray}
	\sum_{m=0}^x n(x,m) = n(x,x) + \sum_{m=0}^{x-1} n(x,m) - \sum_{m=x+1}^N n(m,x),
\end{eqnarray}
where $N$ is the maximum number of pairs in all PNe.
This is a recurrence relation of $x$, to be solved with observational data and simulations together.
Although observational images are available in many PN catalogues, the projection effect makes it hard to tell whether an object with apparently $x$-pairs of lobes should be put in the first or third term on the right hand side of the above equation. 
Therefore at the moment it would be unwise to form a conclusion on the real fraction of multipolar PNe.
On the other hand, it is expected that the fraction of apparent multipolar PNe \[ \left.\sum_{m=2}^{x} n(x,m) \middle/ \sum_{m=0}^{x} n(x,m) \right. \] will increase with the actual number of pairs $x$, and thus the number of nebulae with over three lobes is more closely estimated than the three-paired ones.

\subsection{The Effect of Sensitivity} \label{sen}

\begin{figure}
 \plotone{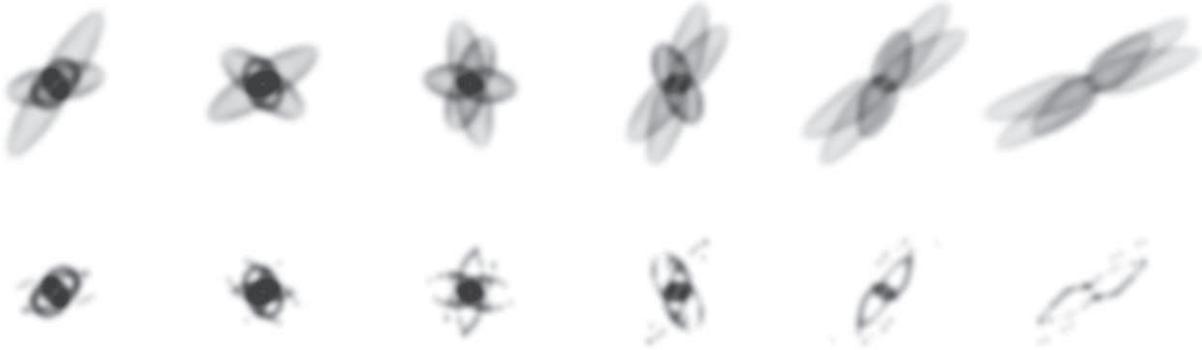} 
 \caption{Perception of morphology is affected by viewing angles and sensitivity. Upper row: with the six angles in the model fixed, the viewing angle is changed from each image to the next by $15^\circ$ of $i$ and $15^\circ$ of PA together. Lower row: each image is modified from the one above so that the faintest pixels below one-third of the peak brightness are cut off. Brightness levels are shown in linear scale.}
   \label{figsen}
\end{figure}

In real observations, the morphological classification also suffers from sensitivity issues.  In images of limited sensitivity or dynamic range, fainter features will be missed and their multipolar nature not recognized.
Figure~\ref{figsen} illustrates a comparison between the images under high and low sensitivity conditions.  The second row images are the same as the top row except that the pixels with brightness less than 1/3 of the peak brightness are not displayed.  If one were to classify the apparent morphology based on the second row images, then one would arrive at very different morphological classes for the objects.

Moreover, the ``broken" segments of faint lobes create an illusion of the presence of minor structures.
The bottom left panel in Figure~\ref{figsen} looks like a bipolar nebula with point-symmetric ansae, filaments or knots around the lobes, similar to the Cat's Eye Nebula \citep[NGC~6543,][]{bal04} and NGC~3242 \citep{rui11}.
In fact, these features are the overlapping areas of two or more lobes:  overlapping increases the column densities of walls so that they can still be seen while the other parts of the lobes are filtered out.

\subsection{Comparison with Observations}

\begin{figure}
 \plotone{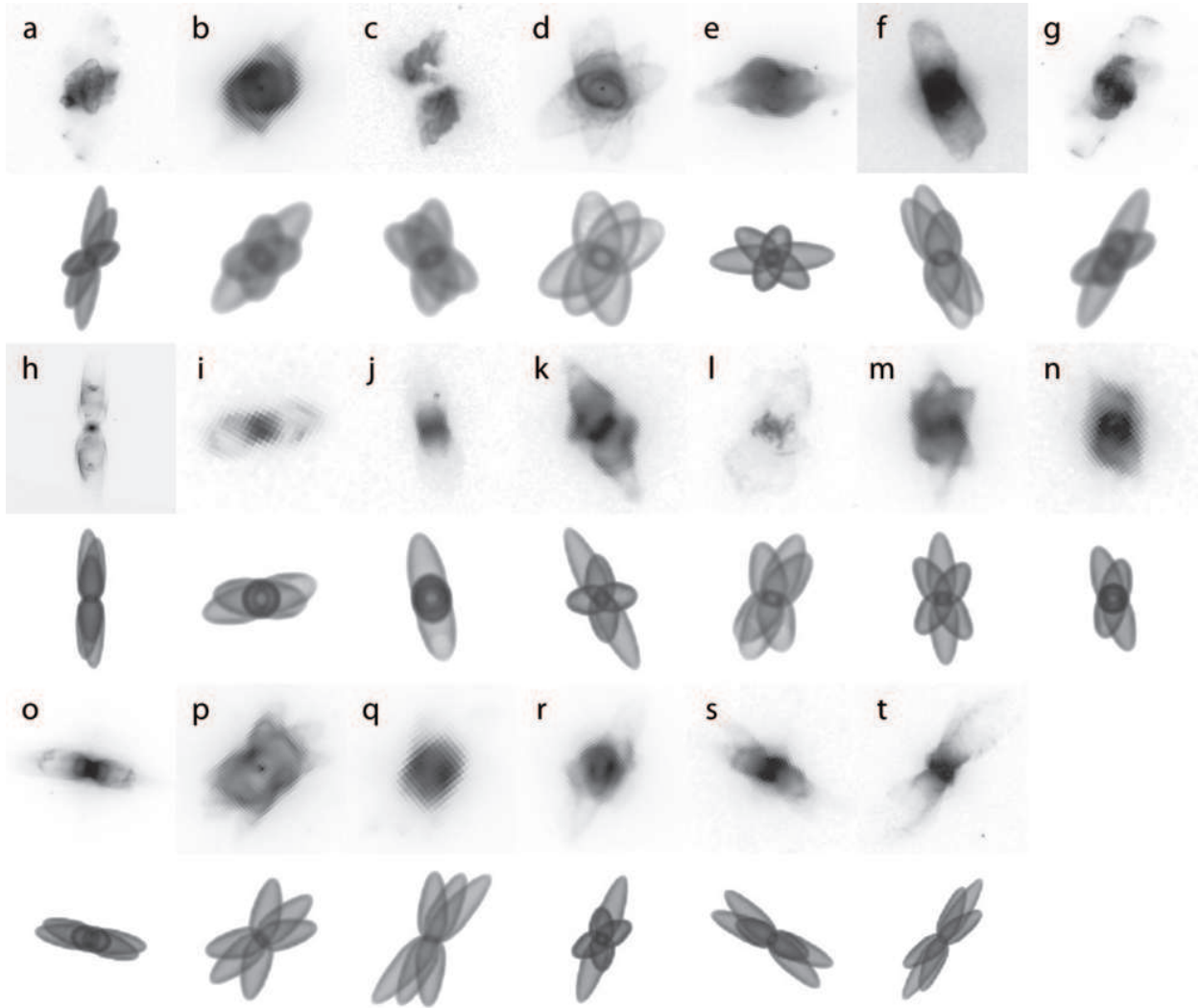} 
 \caption{Comparison of the 20 observed images (those with an index letter), each with its corresponding modeled image (the one right below each observed image). Brightness levels are in log scale. North is pointing up and east to the left. Refer to Table~\ref{tab1} for the object data. The angular sizes and brightness levels of the objects are not the same.}
   \label{hst}
\end{figure}

In this section, we explore whether the simulated images are of any relevance to real observed PNe. We match the 20 selected real PNe images with simulated images of our simple model. 
While we  do not claim that the sample is representative of all multipolar PNe, we hope to demonstrate that our model can be the first-order approximations to the true morphologies of the selected objects.
For simplicity, we change only the six angles as described in Section~\ref{appmor} and keep all of the other parameters fixed.  The effect of sensitivity is not included.
Comparisons between the observed and simulated images are presented in Figure~\ref{hst}.  The parameters used in the simulated images are listed in Table~\ref{tab1}.  The labels of the objects in Figure~\ref{hst} are given in the first column of Table~\ref{tab1}.

\subsection{Individual Objects}

Details of each object in Figure~\ref{hst} are given below.  Here, we try to relate the observed morphological features to the 3D model.

(a) \textit{IRAS 05028+1038} (J~320, PN G190.3-17.7) was classified as a Type II low-excitation PN \citep{har04} with central star effective temperature $T_\textrm{eff} = 85000$K and mass $M_\textrm{CS} =  0.79 M_\sun$ \citep{mcc97}.  It was described as a ``poly-polar" PN with surrounding knots distributed in a point-symmetric pattern \citep{har04}.  The \textit{HST} image clearly shows two knots both to the north and the south, aligned not exactly in the direction of the central parts.
In our model we align the two pairs of lobes in the direction of the knots and align the remaining one to the central lobes in the NW$-$SE direction.
Knots are regarded as higher order properties that can be represented by enhancing densities at the tips, or as results due to sensitivity (see Section~\ref{sen}).
According to \citet{har04}, there is one more knot further away to the west. This may be related to our third pair.

(b) \textit{IRAS 07172-2138} (M1-12, PN G235.3-03.9) has closed collimated pairs of lobes in a point-symmetric shape \citep{sah11}.
The central region is due to two overlapping pairs and the extended features in the NW-SE direction are explained as the third pair.

(c) \textit{IRAS 10197-5750} (Hen 3-404, OH 284.2-0.8, Roberts~22) was described as a bipolar DUPLEX (DUst-Prominent Longitudinally-EXtended) nebula with an A2 I central star by  \citet{uet07}.
Fluorescence-induced H$_2$ emission, characteristic of young PNe, has been detected \citep{gar02}. 
It is also associated with intense OH maser emission \citep{all80}.
The equatorial dark lane in the optical image, together with the near- and far-infrared (IR) image, implies the presence of a dusty ring \citep{all80, cox11}, which is not included in our model.

(d) \textit{IRAS 10214-6017} (Hen 2-47, PN G285.6-02.7) was classified as a young, low-excitation PN with cool IR continua and [Ne {\sc ii}] 12.8~$\mu$m emission \citep{vol90}.
It has six obvious closed lobes which appear to be axis-symmetric about the equatorial direction.
The components in the east side have a larger size than the others, and \citet{sah00} further divided them into two more lobes.
The central bright ring \citep{sah00} is reproduced in the modeled image as the overlapping regions of the three pairs, assuming a simple addition of the column density.

(e) \textit{IRAS 15015-5459} (Hen 2-115, PN G321.3+02.8) has a central star of $T_\textrm{eff} = 27400$K and $M_\textrm{CS}=0.68 M_\sun$ \citep{zha93} surrounded by warm dust.
In addition to a pair of elongated lobes, there are point-symmetric bulges nearer to the center.
\citet{sah98} identified a small rhomboidal structure around the central star, which may also be an illusion due to overlapping of the multiple pairs in our model.

(f) \textit{IRAS 16409-1851} (Hen 2-180, PN G000.1+17.2) shows two lobes slightly bent to the west. Such ``bending" structures are very difficult to understand by traditional physical ejection models, but can arise naturally in this simulation: the bent lobes seen at the west side belong to two pairs of symmetric lobes but their counterparts in the east are significantly fainter. It has a narrow bright waist \citep{sah11}, which is again a possible consequence of superposition of multiple lobes.

(g) \textit{IRAS 16585-2145} (IC 4634, PN G000.3+12.2) has a curious shape consisting of an S-shape main nebula with extended S-shape bow-shock structures studied by \citet{gue08}.
In the central region, an inner shell is hidden roughly aligned with the main axis and knots are found along the bow-shock features \citep{gue08}.
According to \citet{sch93a}, the central part exhibits both red and blue shifted features on the same side along the main axis, suggesting the presence of a fast precessing collimated outflow with time-dependent velocity for shaping the nebula.
The S-shape point-symmetric structure of the nebula can be explained by two overlapping bipolar lobes and 
the observed S-shape can be reproduced  by non-uniform density distribution along the azimuthal angle of the lobes.

(h) \textit{IRAS 17028-1004} (Butterfly Nebula, M~2-9, PN G010.8+18.0) has highly collimated open-ended lobes with skirt-like structures \citep{sah11}.  
Bright knots, or ansae, are found along the long axis, possibly formed at the head of an early jet \citep{sok90}.
The proper motions of the dusty blobs much further away from the main nebula are studied by \citet{cor11}. Their work postulated a rotating ionizing beam to explain the ionized gas emission phenomenon in this nebula, and supported the idea that the appearance of multiple lobes was the result of an excitation gradient.
However, considering that other nebulae with similar apparent morphology may not have been studied in as much detail as M~2-9 to reveal the physical properties, we present the model here to suggest an alternate possible choice.
The modeled image suggests that the fork-like ends of the main lobe can be geometrically reproduced by two pairs of lobes, and the ansae (knots) are at the interacting points of the three pairs.
There are fine structures such as arcs not included here. In addition, the lobes are more likely  to be open-ended (see Section~\ref{lobeshp}).

(i) \textit{IRAS 17156-3135} (PN G354.5+03.3) In low dynamic range images, the nebula has the typical shape of a bipolar nebula. The \textit{HST} image reveals another pair of bipolar lobes with an axis aligned at an angle with respect to the primary bipolar lobes.
We put the third pair of bipolar lobes to be pole-on in order to produce the circular feature at the center. Given three pairs of lobes, it would not be uncommon that one pair would lie close to perpendicular to the plane of the sky.  

(j) \textit{IRAS 17296-3641} (PN G351.9-01.9) has an inner bubble inside the main bipolar lobe pair \citep{sah11}.
Considering the brightness difference between the central part and the lobes, we put two pairs as nearly pole-on.

(k) \textit{IRAS 17389-2409} (Hen 2-267, M 2-14, PN G003.6+03.1) is a point-symmetric PN with bulges.
The apparent shape is similar to IRAS~15015-5459, but with a bright waist that is not perpendicular to the largest pair of lobes.

(l) \textit{IRAS 17410-3405} (Hen 2-271, M 3-14, PN G355.4-02.4) is an asymmetric PN with apparently more bubbles in the S$-$E side than the N$-$W side.
The H$\alpha$ brightness peaks at the central region and drops slowly outward.
The NW bubbles are fainter and not as clearly seen as the SE ones. This asymmetry can be due to non-uniform brightness distributions on the two sides, or can be the result of circumstellar dust extinction. 
Besides the three pairs, there is another small bubble in the equatorial direction.

(m) \textit{IRAS 17496-2221} (Hen 2-299, M 1-31, PN G006.4+02.0) 
has a peculiar hexagonal shape that can be reproduced by three pairs of bipolar lobes, with a pair near the plane of the sky representing the long axis of the nebula.  The bright waist represents the overlapping central parts of the three pairs of lobes.

(n) \textit{IRAS 17549-3347} (Hen 2-313, PN G357.1-04.7) 
The central part of this nebula has a circular shape which is considered as a pair of nearly pole-on lobes.
The more extended structure has a pointed S-shape.

(o) \textit{IRAS 17567-3849} (Hen 2-320, PN G352.9-07.5) 
has a brighter, shorter pair of lobes and two narrower, longer pairs described as ``shoulders" by \citet{den09}. These authors suggest that this structure can be produced by bow shocks in their magnetohydrodynamic nested-wind simulations.
More extended collimated ``nose cones" have also been found.  Such double ``inner'' and ``outer'' bipolar lobes can be naturally explained by a multipolar structure as seen in the simulated image.

(p) \textit{IRAS 18022-2822} (Hen 2-339, M 1-37, PN G002.6-03.4) 
has three pairs of lobes distributed in a point-symmetric manner.
An arc-like feature to the north is believed to be part of a halo \citep{sah11}.
The central part seems to be hollow and a bright star can be seen.  This classic multipolar structure can be reproduced by three pairs of bipolar lobes almost on the plane of the sky.

(q) \textit{IRAS 18039-2913} (Hen 2-346, PN G002.1-04.2) exhibits faint extensions around a bright irregular center.  The close-ended barrel recorded by \citet{sah11} probably refers to the center.
Three bullet-like features can be seen in the SW part and there are faint lobes in the opposite direction.
The three lobes in our model image point to these features but fail to reproduce the central barrel.
An extra component may be needed.

(r) \textit{IRAS 18430-1430} (M 1-61, PN G019.4-05.3) 
has one pair of narrow lobes stretching out from the point-symmetric central part which is possibly another two pairs of lobes.

(s) \textit{IRAS 19431+2112} (Hen 2-447, PN G057.9-01.5) has a bright waist and point-symmetric lobes.
The shape of two lobes are more obvious in the NE direction.
Knots and filaments were identified \citep{cui05}.

(t) \textit{IRAS 20090+3715} (Hen 2-456, NGC 6881, PN G074.5+02.1) has a complicated morphology.
\citet{kwo05} found four pairs of rings in the multipolar lobes aligned with one pair of lobes, and three rings at the waist aligned to another lobe pair.
\citet{gue98} identified a loop to the SE of the main nebula, almost perpendicular to the major axis.
The H$_2$ image shows a much more extended size than the optical one, and the lobe to the NW is even larger than its counterpart, turning into an irregular shape \citep{ram08}.
On the other hand, the 6~cm radio morphology is more elongated in the equatorial direction \citep{aaq90}, and only images the region with the highest emission measure.
Our model can simulate the multipolar lobes but not the central bright region.

\section{Discussions} \label{discuss} 

PNe have many different morphological features  which are often attributed to different physical regions created by separate physical processes. Our models suggest that many of the observed features can be explained by a single, unified model. Of the bright torus seen in many PNe, some are due to real volume density enhancements, but some probably represent the overlapping region of different lobes. Simulations similar to the one presented here can sort out the ``false torus" percentage in the whole PN population. The point-symmetric S-shape morphology can be explained by nearly aligned pairs of lobes. Ansae may be the tips of unseen lobes. The double inner$-$outer bipolar lobes can arise from overlapping multipolar lobes. Due to the long path length, a pair of near pole-on bipolar lobes can easily be mistaken for a torus, as in the case of NGC 6720 \citep{kwo08}. The observed images of PNe can be successfully modeled with a very simple multipolar model.  The model lobes have no density variation (either radial or azimuthal) and all three lobes have identical shapes and sizes. There is no inclusion of the effects from dust extinction, which can make the front lobes brighter and the back lobes fainter.

\subsection{The Shape of the Lobes}\label{lobeshp}


The shapes of the lobes reflect the dynamical history of the outflows and the interactions among outflows.  Assuming a collimated fast outflow from the central star, the shapes of the lobes are defined by the interaction between this fast outflow and the ambient circumstellar materials, e.g. the remnant of the AGB wind.  At a later stage of development, the fast outflow may break out of the circumstellar envelope and change into an open-ended shape.

\begin{figure}
 \plotone{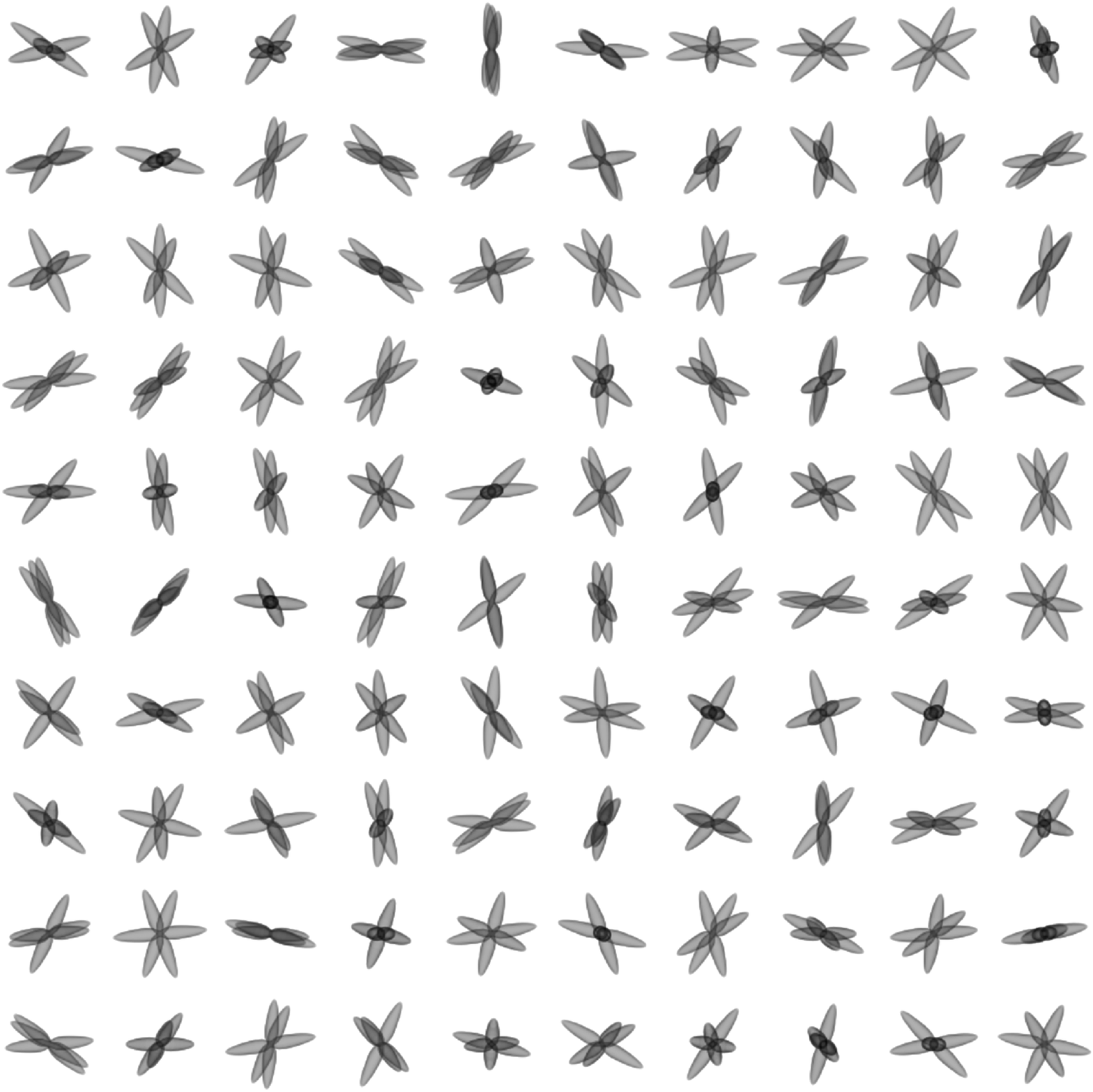}
 \caption{100 projected images of the three-lobed model with randomly generated inclination and position angles for each pair. The scales are the same in all panels.}
 \label{figrdm}
\end{figure}

\begin{figure}
 \plotone{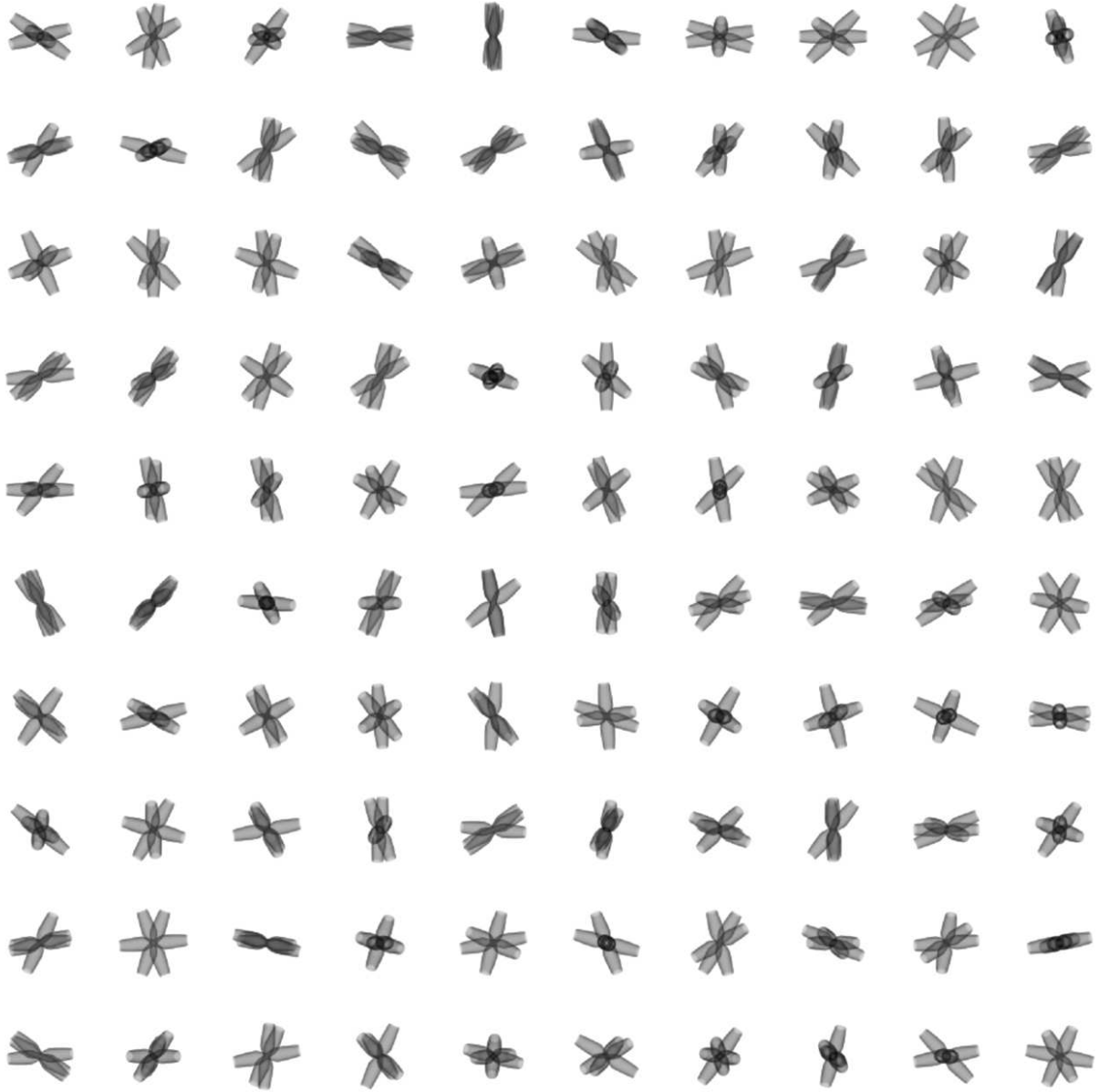}
 \caption{Same as Figure~\ref{figrdm} but with open-ended lobes.}
 \label{opnrdm}
\end{figure}

\begin{figure}
 \plotone{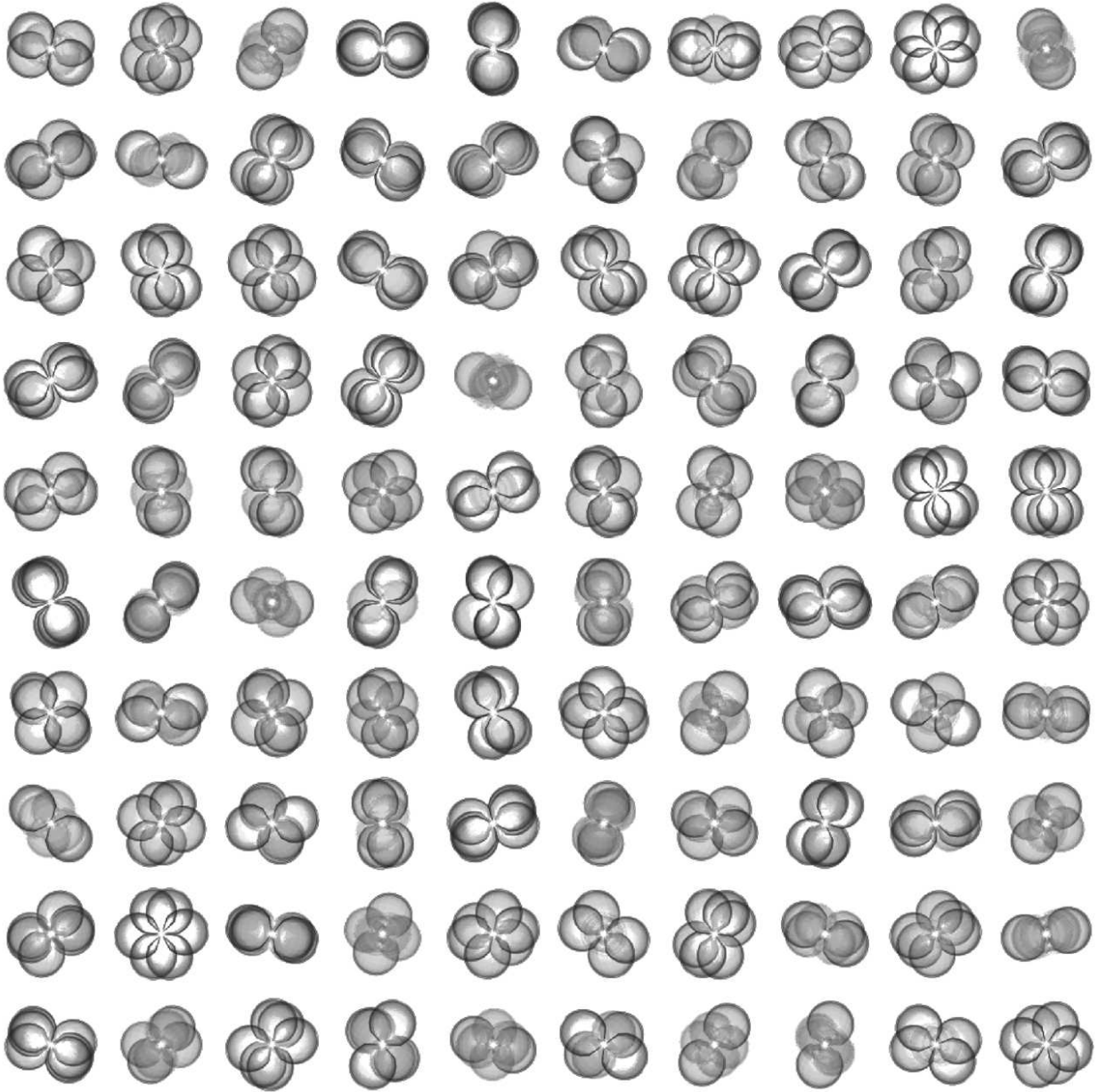}
 \caption{Same as Figure~\ref{figrdm} but with a different lobe shape. See the text for the descriptions.}
 \label{rdm2}
\end{figure}

The lobes are set to be close-ended in our model.  The difference between close- and open-ended lobes is easily identified even in a projected image, unless the lobe is tilted almost pole-on.
This can be seen by comparing Figure~\ref{figrdm} and Figure~\ref{opnrdm}.
Moreover, the identification of multiple lobes does not depend on this factor.  Figure~\ref{rdm2} shows another example of the closed-lobed model with a different lobe shape for comparison, described by the equation $r=\cos\theta$ in spherical coordinates.  Compared to Figure~\ref{figrdm}, Figure~\ref{rdm2} shows greater deviations from the real PN images.
In fact, lobes with larger opening angles have a higher chance of interacting with each other and producing more complicated shapes.

\subsection{The Number of Pairs}

In general, ``multipolar" means having more than one pair of lobes, and is not confined to objects with three pairs.
For the objects chosen, at least three pairs are obviously seen, and the number three is also commonly found in the literature (e.g., NGC~7027 by \citealt{nak10}; NGC~6644 by \citealt{hsi10}; and NGC~7026 by \citealt{cla12}). 
It is possible that there are more than three pairs (e.g. IRAS~19024+0044 by \citealt{sah05}; and NGC~5189 by \citealt{sab12}), but adding more pairs means adding more parameters; at this stage we hope to keep the number of parameters to a minimum.
The fourth pair is usually not as obvious as the other three.
For IRAS~19024+0044, the fourth pair almost overlaps with another one, which could be due to projection effects, or the two components may have been in the same lobe produced by perturbations.
At least three is enough to describe such objects.
The less obvious lobes are therefore treated as higher ordered structures.

\subsection{Lobes: Same Length or Not?}
By looking at a 2D image, it is hard to tell whether the lobes have the same length or not without knowing the inclination angles.
One of the possible ways to predict the length ratios is to compare the brightness of various lobes.
If the brightness is simply proportional to the square of the column density, and if the volume density is constant everywhere, then the lengths along the LOS can be calculated.
As shown in the modeled images, lobes that are more pole-on look brighter.
However, infrared images confirmed the existence of denser tori in some pPNe, e.g., IRAS~16594-4656 \citep{vol06} and IRAS~17441-2411 \citep{vol07}, which may be optically thick in the visible light.
Even if the central bright ionized region is optically thin in PNe, it is still hard to tell whether it is a dense, short component or a less dense but longer one.

Another way to determine the actual length ratios is from kinematic information.
Lobes with the same lengths are likely to be produced simultaneously and with the same outflow velocities.
Assuming constant expansion velocities, the ratio of the apparent size to the velocity component along the LOS should be able to indicate the projected angles of each lobe.
Take $i$ to be the angle between the tilted lobe and the LOS.
The apparent angular size $r_\textrm{app}$ is related to the real size $r_\textrm{real}$ by a factor of $\sin i$, while the LOS velocity and real velocity are linked by $\cos i$:
\begin{eqnarray}
	r_\textrm{app} &=& r_\textrm{real} \sin i,	\\
	v_\textrm{LOS} &=& v_\textrm{real} \cos i.
\end{eqnarray}
To find $r_\textrm{real}$ and $v_\textrm{real}$ one must know $i$.
For one single pair of lobes, it is hard to deduce $i$, but with multiple lobes, the ratios between two pairs can eliminate the real quantities, based on the assumption of the same $r_\textrm{real}$ and the same $v_\textrm{real}$ for all lobes.
On the other hand, for lobes of different lengths, deducing the true length ratios from 2D images is more uncertain.

A detailed morpho-dynamical study is the key to a reliable determination of the real shape of a PN. The kinematic information is the only way to ``see" the third dimension in the 2D sky plane. In addition to the estimation of the length of the lobes, visually overlapping lobes can also be distinguished from their radial velocities because they are tilted with different angles to the LOS.

\subsection{Formation and Evolution of Multipolar Lobes}
Although the formation mechanism of multipolar lobes is not confirmed, the theory of a precessing jet is more frequently adapted than coeval mass outflows.
As in the case of bipolar PNe, a binary system is believed to be a driving source.
The jet changes direction due to the additional velocity component of the progenitor's orbital motion \citep[][and references therein]{vel12} and describes a spiral trajectory, which can be traced by motions of water maser spots in an earlier phase \citep{ima02}.
It has been suggested that the number of lobes is proportional to the ratio of the jet precession period to the orbital period, which depends on the masses of the two objects in the binary system \citep{vel12}.
In this hypothesis, the angles between adjacent lobes should be equal, and all the lobes should be confined to a cone swept out by the jet.
Compared with our random direction simulation, this hypothesis puts more constraints on the lobe orientations, and therefore projection effects are expected to be more significant: the lobes have a smaller angular distance to each other and thus a higher chance of overlap in the projected images.
Furthermore, if the precessing jet mechanism is true, then the S-shape structure found in some PNe is just a variation of the multipolar scenario, where the ``lobes" are so close to each other that a continuous S-shape is seen instead of discrete lobes.

In each lobe, expansion components in the direction perpendicular to the lobe axis push out the lobe walls.
Eventually, the lobes interact with each other, combine into a larger lobe, or become less obvious because of the fusion of the walls.
As a result, the phenomenon of multipolar lobes is more commonly observed in pPNe or young PNe.
If the existence of a companion is the only way to produce multipolar outflows, then the total fraction of multipolar PNe can be more easily estimated as the product of the fraction of binary systems in all PN progenitors and the fraction of multipolar outflows onset in these binary systems, scaled by the ratio of survival duration of multiple lobes to the lifetime of a PN.
Although the fraction of binary systems directly confirmed in individual PN progenitors is lower than the percentage of multipolar PNe (12\%$-$20\%, see the introduction section), the overall estimated fraction of binary progenitors is comparable: the close binary fraction might be as high as 12\%$-$21\% \citep{mis09} or $22\%\pm9$\% \citep{fre10}.


\section{Summary}

From a simple three-bipolar-lobes model, we have shown that many of the observed morphological features of PNe can be successfully reproduced.  Morphological features such as ``point-symmetric'', ``S-shape'', ``bending'', ``double bipolar lobes'', ``knots'', and ``ansae'' can arise naturally from this model.  These simulations confirm the importance of the effects of orientation and sensitivity, as discussed in \citet{kwok2010}.  We have also shown that the statistical analysis of morphological classes based on apparent shapes is not a reliable indicator of the true structures of the nebulae.

The reconstruction of the true 3D structure of PNe represents the first step in the identification of the physical processes responsible for the shaping of PNe.  Without an accurate  picture of the 3D structure, all attempts at physical models are futile.  

Further expansion of the model should take into account the density profiles, ionization and  radiative transfer treatments, dust extinction, and  interaction with the circumstellar and interstellar media. The fourth dimension $-$ temporal evolution is just as important.
The 20 PNe presented are barely the tip of the iceberg among the known multipolar PNe, and our simulations suggest that the number of multipolar PNe will keep on rising as the sensitivity and dynamic range of images improve with future observations.


\acknowledgments

We thank Nico Koning and Wolfgang Steffen for discussions on the use of the SHAPE software.  We thank Romano Corradi for useful ideas and comments. The work was supported by a grant awarded to SK from the Research Grants Council of the Hong Kong Special Administrative Region, China (Project No. HKU 7031/10P).




\clearpage

\clearpage

\begin{deluxetable}{crrrrrrc}
  \tablecaption{Randomly Generated Angles of Each Image in Figure~\ref{figrdm}, from Left to Right, from Top to Bottom.}
  \tablehead{\colhead{} & \colhead{$i^\circ_1$} &\colhead{PA$^\circ_1$} & \colhead{$i^\circ_2$} & \colhead{ PA$^\circ_2$} & \colhead{$i^\circ_3$} & \colhead{PA$^\circ_3$} & \colhead{Class}}
  \startdata
1 & 83 & -125 & 64 & 105 & 22 & -120 & 2\\
2 & 72 & 170 & 57 & 120 & 61 & 25 & 3\\
3 & 26 & 105 & 22 & -130 & 65 & -35 & 3\\
4 & 66 & 95 & 76 & 75 & 78 & 100 & 2\\
5 & 84 & -175 & 71 & 0 & 55 & -10 & 2\\
6 & 35 & -130 & 85 & -95 & 40 & 50 & 2\\
7 & 85 & 70 & 29 & 175 & 89 & 100 & 3\\
8 & 76 & 135 & 42 & 50 & 81 & 90 & 3\\
9 & 80 & -130 & 84 & -25 & 90 & -65 & 3\\
10 & 16 & -75 & 29 & -175 & 50 & 20 & 2\\
11 & 74 & 105 & 54 & -30 & 51 & -80 & 2\\
12 & 27 & 115 & 14 & 105 & 79 & 75 & 2\\
13 & 61 & -20 & 73 & -55 & 68 & 175 & 3\\
14 & 61 & 65 & 90 & 40 & 49 & -105 & 3\\
15 & 56 & -50 & 49 & 145 & 79 & -70 & 3\\
16 & 73 & 25 & 47 & 100 & 60 & 20 & 2\\
17 & 27 & 125 & 68 & -40 & 42 & 180 & 2\\
18 & 56 & 170 & 82 & -135 & 26 & 20 & 2\\
19 & 58 & 130 & 37 & -15 & 71 & -175 & 3\\
20 & 64 & -85 & 85 & 125 & 58 & 135 & 3\\
21 & 34 & -45 & 80 & -155 & 69 & 110 & 3\\
22 & 89 & 0 & 49 & 155 & 85 & 45 & 3\\
23 & 69 & 70 & 75 & -170 & 58 & -20 & 3\\
24 & 30 & -115 & 79 & -135 & 85 & 65 & 2\\
25 & 67 & -60 & 69 & -75 & 46 & -165 & 3\\
26 & 52 & 170 & 87 & -125 & 80 & 25 & 3\\
27 & 78 & -170 & 66 & 160 & 90 & -70 & 3\\
28 & 62 & -30 & 83 & 105 & 63 & -35 & 2\\
29 & 51 & 180 & 88 & 150 & 41 & 60 & 3\\
30 & 89 & -35 & 65 & 145 & 78 & -5 & 2\\
31 & 56 & 140 & 87 & 95 & 84 & -55 & 3\\
32 & 63 & -45 & 38 & 115 & 69 & 150 & 3\\
33 & 63 & -45 & 58 & 55 & 90 & -15 & 3\\
34 & 88 & -30 & 64 & 120 & 78 & -10 & 3\\
35 & 43 & -110 & 16 & 150 & 11 & -70 & 1\\
36 & 22 & -30 & 55 & -130 & 90 & 180 & 2\\
37 & 44 & 45 & 56 & -110 & 77 & 15 & 3\\
38 & 30 & 115 & 65 & -10 & 88 & 165 & 2\\
39 & 37 & -170 & 75 & 105 & 64 & -160 & 2\\
40 & 63 & 110 & 87 & 60 & 57 & 60 & 2\\
41 & 63 & 140 & 26 & -95 & 82 & -90 & 2\\
42 & 75 & -165 & 55 & 175 & 21 & -80 & 2\\
43 & 80 & 20 & 30 & 140 & 61 & 180 & 3\\
44 & 65 & -30 & 51 & -115 & 43 & -155 & 3\\
45 & 11 & 105 & 88 & 135 & 81 & 95 & 2\\
46 & 84 & 15 & 58 & 30 & 52 & -50 & 3\\
47 & 8 & -5 & 71 & 10 & 60 & -45 & 2\\
48 & 52 & 60 & 42 & -155 & 48 & 115 & 3\\
49 & 84 & 30 & 90 & 160 & 83 & -125 & 3\\
50 & 76 & 150 & 84 & -145 & 72 & -170 & 3\\
51 & 81 & 15 & 80 & 40 & 81 & 25 & 3\\
52 & 55 & -45 & 70 & 140 & 31 & -55 & 1\\
53 & 34 & -155 & 3 & -95 & 51 & 85 & 2\\
54 & 31 & 90 & 90 & -15 & 82 & 150 & 3\\
55 & 73 & -170 & 90 & 10 & 70 & -50 & 2\\
56 & 62 & 10 & 59 & -10 & 29 & -145 & 3\\
57 & 63 & -75 & 73 & -40 & 39 & -105 & 3\\
58 & 71 & 90 & 82 & 80 & 60 & -50 & 3\\
59 & 44 & 100 & 20 & -125 & 90 & 125 & 2\\
60 & 75 & 30 & 68 & 150 & 55 & -100 & 3\\
61 & 70 & 145 & 83 & -135 & 51 & 55 & 2\\
62 & 52 & 60 & 78 & -75 & 23 & 65 & 2\\
63 & 66 & 15 & 57 & -145 & 57 & -60 & 3\\
64 & 39 & -55 & 67 & -175 & 55 & 40 & 3\\
65 & 85 & -140 & 48 & -145 & 89 & 175 & 2\\
66 & 74 & -110 & 77 & 5 & 55 & 100 & 3\\
67 & 12 & -100 & 39 & 55 & 65 & 150 & 2\\
68 & 58 & 110 & 22 & -55 & 70 & -165 & 2\\
69 & 9 & -70 & 89 & 65 & 48 & 160 & 2\\
70 & 13 & 5 & 60 & 100 & 61 & -110 & 2\\
71 & 30 & 60 & 29 & -10 & 86 & 50 & 2\\
72 & 66 & -105 & 51 & 5 & 75 & -30 & 3\\
73 & 44 & -145 & 54 & -155 & 82 & 100 & 2\\
74 & 69 & -160 & 20 & -35 & 70 & -5 & 2\\
75 & 72 & 120 & 88 & -50 & 71 & 85 & 3\\
76 & 39 & -35 & 32 & 160 & 54 & -10 & 2\\
77 & 37 & 75 & 54 & 135 & 67 & -115 & 2\\
78 & 55 & 5 & 64 & 0 & 86 & -40 & 2\\
79 & 37 & -110 & 83 & -65 & 69 & -90 & 3\\
80 & 62 & -35 & 16 & -5 & 40 & -110 & 2\\
81 & 63 & -75 & 67 & 155 & 58 & 95 & 2\\
82 & 82 & -20 & 86 & 95 & 90 & 20 & 3\\
83 & 50 & 75 & 55 & -95 & 89 & -105 & 1\\
84 & 40 & -105 & 52 & -20 & 15 & -90 & 2\\
85 & 68 & 165 & 53 & -115 & 76 & 100 & 3\\
86 & 76 & -105 & 6 & 65 & 67 & 20 & 2\\
87 & 81 & -30 & 72 & -165 & 61 & -60 & 3\\
88 & 70 & -115 & 44 & 90 & 28 & -135 & 2\\
89 & 69 & -85 & 59 & -10 & 54 & 125 & 3\\
90 & 68 & -80 & 30 & -75 & 13 & -80 & 1\\
91 & 78 & 50 & 71 & 60 & 71 & 95 & 2\\
92 & 32 & -25 & 53 & -35 & 46 & 90 & 2\\
93 & 73 & -5 & 64 & 155 & 77 & -80 & 3\\
94 & 82 & 155 & 44 & -145 & 54 & -135 & 2\\
95 & 35 & 95 & 58 & 75 & 33 & -5 & 3\\
96 & 36 & -40 & 90 & -120 & 53 & 120 & 3\\
97 & 67 & 145 & 29 & 80 & 27 & -165 & 3\\
98 & 17 & 55 & 27 & -140 & 62 & 165 & 1\\
99 & 72 & -80 & 12 & 85 & 82 & 45 & 2\\
100 & 90 & 80 & 66 & -25 & 52 & 30 & 3\\
  \enddata
  \tablecomments{See Section~\ref{appmor} for the descriptions on ``Class" and other details.}
  \label{tblrdm}
\end{deluxetable}

\begin{deluxetable}{cccrrrrrrccc}
  \tablecaption{Information and parameters of the 20 PNe.}
 \tablehead{
 {} & \colhead{IRAS Name} & \colhead{Dataset}  & \colhead{$i^\circ_1$} &\colhead{PA$^\circ_1$} & \colhead{$i^\circ_2$} & \colhead{PA$^\circ_2$} & \colhead{$i^\circ_3$} & \colhead{PA$^\circ_3$} &  \colhead{$\theta^\circ_{min}$} & \colhead{$\theta^\circ_{med}$} &  \colhead{$\theta^\circ_{max}$}  
 }
\startdata
a & 05028+1038 & U39H1301B & 90 & -8 & 48 & -25 & 22 & -64 & 33.1 & 44.8 & 77.9 \\ 
 b & 07172-2138 & U5HH0502B & 16 & -75 & 17 & -10 & 33 & -41 & 17.6 & 20.0 & 21.7 \\ 
 c & 10197-5750 & U3B30201B & 26 & -2 & 20 & 24 & 23 & 57 & 12.0 & 12.0 & 23.7 \\ 
 d & 10214-6017 & U35T1407B & 22 & 11 & 23 & -62 & 21 & -30 & 12.1 & 14.6 & 26.3 \\ 
 e & 15015-5459 & U35T2905B & 37 & 87 & 17 & -21 & 21 & 40 & 19.0 & 33.5 & 51.7 \\ 
 f & 16409-1851 & U5HH3102B & 20 & 0 & 36 & 17 & 32 & 33 & 9.8 & 18.0 & 18.7 \\ 
 g & 16585-2145 & U47B0201B & 48 & -25 & 25 & -63 & 14 & -25 & 16.0 & 31.2 & 33.4 \\ 
 h & 17028-1004 & U42I0202B & 30 & 0 & 55 & 2 & 45 & -6 & 11.7 & 15.5 & 25.1 \\ 
 i & 17156-3135 & U6MG5001B & 20 & 85 & 27 & -74 & 0 & 0 & 11.0 & 20.1 & 27.1 \\ 
 j & 17296-3641 & U6MG4801B & 3 & 18 & 31 & 12 & 6 & 9 & 3.0 & 24.5 & 27.4 \\ 
 k & 17389-2409 & U6MG1501B & 50 & 26 & 18 & -84 & 24 & 7 & 28.2 & 29.6 & 46.2 \\ 
 l & 17410-3405 & U6MG3101B & 25 & -40 & 32 & -23 & 25 & 12 & 10.5 & 17.7 & 21.6 \\ 
 m & 17496-2221 & U5HH6902B & 40 & 3 & 23 & -35 & 27 & 33 & 20.6 & 25.7 & 27.3 \\ 
 n & 17549-3347 & U6MG3601B & 0 & 0 & 31 & 18 & 22 & -14 & 16.6 & 21.6 & 31.2 \\ 
 o & 17567-3849 & U5HH1302B & 41 & 82 & 35 & 71 & 10 & 79 & 8.5 & 25.1 & 30.3 \\ 
 p & 18022-2822 & U35T2105B & 37 & -12 & 39 & -50 & 33 & -77 & 17.5 & 23.3 & 36.1 \\ 
 q & 18039-2913 & U5HH4103B & 41 & -3 & 45 & -26 & 58 & -42 & 16.0 & 18.4 & 34.0 \\ 
 r & 18430-1430 & U59B0301B & 21 & -62 & 52 & -19 & 20 & 0 & 20.5 & 33.2 & 38.4 \\ 
 s & 19431+2112 & U59B0704B & 51 & 44 & 50 & 67 & 27 & 65 & 17.8 & 23.5 & 27.7 \\ 
 t & 20090+3715 & U39H3601B & 50 & -22 & 40 & -55 & 80 & -33 & 25.1 & 31.6 & 43.9 \\  
\enddata
\tablecomments{The order of pairs 1, 2 and 3 is arbitrary. $\theta$ (minimum, median or maximum) refer to the separation angles between pairs. }
\label{tab1}
\end{deluxetable}

\end{document}